\documentclass[12pt]{article}
\usepackage{epsfig}
\usepackage{color}
\usepackage{amssymb,amsmath}

\newcommand{\bea}{\begin{eqnarray}}
\newcommand{\eea}{\end{eqnarray}}
\newcommand{\be}{\begin{equation}}
\newcommand{\ee}{\end{equation}}

 \makeatletter
\def\alt{\mathrel{\mathpalette\gl@align<}}
\def\agt{\mathrel{\mathpalette\gl@align>}}
\def\gl@align#1#2{\lower.6ex\vbox{\baselineskip\z@skip\lineskip\z@
\ialign{$\m@th#1\hfil##\hfil$\crcr#2\crcr\sim\crcr}}} \makeatother

\begin{document}
\begin{flushright}
\end{flushright}
\vspace*{1.0cm}

\begin{center}
\baselineskip 20pt
{\noindent}{\Large\bf
Leptonic $CP$ Violation and Leptogenesis \\
in Minimal Supersymmetric SU(4)$_c  \times$SU(2)$_L \times$SU(2)$_R$
}
\vspace{1cm}

{\large
Nobuchika Okada$^{~a}$ 
 and
Qaisar Shafi$^{~b}$
} \vspace{.5cm}

{\baselineskip 20pt \it
$^{a}$ Department of Physics and Astronomy,
University of Alabama, Tuscaloosa,  AL 35487, USA \\
\vspace{3mm}
$^{b}$ Bartol Research Institute, Department of Physics and Astronomy, \\
University of Delaware, Newark, DE 19716, USA
}

\vspace{.5cm}

\vspace{1.5cm} {\bf Abstract}\\
\end{center}

We consider a supersymmetric SU(4)$_c \times$SU(2)$_L \times$SU(2)$_R$
  model with a minimal number of Higgs multiplets and Dirac and Majorana $CP$-violating phases 
  in the neutrino flavor mixing matrix.
The model incorporates the charged fermion masses and quark mixings,
  and uses type I seesaw to explain the solar and atmospheric neutrino oscillations.  
With the neutrino oscillation data of two mass squared differences and three flavor mixing angles, 
  we employ thermal leptogenesis and the observed baryon asymmetry to find the allowed regions 
  for the Dirac and Majorana phases. 
For a normal neutrino mass hierarchy, we find that the observed baryon asymmetry can be reproduced 
  by  a Dirac phase of around $\delta_{CP}=3 \pi/2$,
  which is strongly indicated by the recent T2K and NO$\nu$A data. 
For the case of inverted neutrino mass hierarchy, the predicted baryon asymmetry is not compatible
  with the observed value.

\thispagestyle{empty}

\newpage

\addtocounter{page}{-1}
\setcounter{footnote}{0}
\baselineskip 18pt

The neutrino oscillation phenomena have established non-zero neutrino masses and mixings
  between different neutrino flavors.
Two neutrino mass squared differences and three mixing angles 
  in the Pontecorvo-Maki-Nakagawa-Sakata (PMNS) mixing matrix
  are measured with good accuracy.
The neutrino oscillation parameters to be determined in future experiments include
  the Dirac $CP$-violating phase ($\delta_{CP}$) in the PMNS mixing matrix 
  and the ordering of the neutrino mass eigenvalues. 
The recent results by the T2K experiment \cite{T2K} strongly indicate a $CP$-violation in the lepton sector 
  with the Dirac $CP$-violating phase of  around $\delta_{CP}=\frac{3 \pi}{2}$.  
The T2K results are also consistent with the results by the NO$\nu$A experiment \cite{NOVA}.
In the not too distant future, the accuracy of measurements for the $CP$-violating phase 
  will be significantly improved. 
A precise information of quark and lepton mass matrices could provide important clues 
  regarding the origin of fermion masses, flavor mixings and $CP$-violations 
  which, most likely, comes from new physics beyond the Standard Model (SM).

In order to explain the observed neutrino masses and flavor mixings,
 we need to extend the SM.
The type I seesaw mechanism \cite{Seesaw}
 is one of the promising ways not only to incorporate the neutrino masses and flavor mixings
 but to also explain the tiny of neutrino masses naturally.
A class of supersymmetric (SUSY) grand unified theories (GUT)
 has attracted much interest in this regard.
In addition to providing a resolution of the gauge hierarchy problem,
 the paradigm of SUSY grand unification is also supported
 by the successful unification of the three SM gauge couplings
 at the GUT scale, $M_{GUT} \simeq 2 \times 10^{16}$ GeV.
Among several possibilities, SO(10) unification
 is one of the more compelling ones, with
 the quark and lepton multiplets of each generation unified
 in a ${\bf 16}$ dimensional representation along with
 a right-handed neutrino.
The seesaw mechanism is also automatically implemented,
 being associated with the breaking of SO(10) symmetry
 to the SM gauge group at $M_{GUT}$, which is fairly
 close to the desired seesaw scale.

The so-called minimal SUSY SO(10) model \cite{BM}
 with the minimal set of Higgs multiplets
 (${\bf 10}$+$\overline{\bf 126}$)
 relevant for fermion mass matrices is a natural extension
 of non-supersymmetric SO(10) models considered
 a long time ago \cite{Lazarides:1980nt}.
Because of the unification of quarks and leptons
 in the ${\bf 16}$ representation and
 the minimal set of Higgs multiplets,
 the fermion Yukawa matrices are highly constrained
 with the quark and lepton mass matrices related to each other.
Note that the Higgs ${\bf 10}$-plet has been used to implement
 $t$-$b$-$\tau$ Yukawa unification in SO(10)
 \cite{Ananthanarayan:1991xp}.
There have been several efforts within the SO(10) framework
 to simultaneously reproduce the observed quark-lepton mass matrix data
 as well as the neutrino oscillation data
 \cite{Fukuyama-Okada, datafit, Mimura}.
It is quite interesting that after the data fitting,
 essentially no free parameter is left and all fermion Yukawa matrices,
 in particular, the neutrino Dirac Yukawa matrix, are unambiguously determined.
The neutrino Dirac Yukawa matrix allows us to provide concrete predictions
 for proton lifetime \cite{p-dcay} and the rate of lepton flavor violations \cite{LFV}.

However, the minimal SO(10) model suffers from a serious problem.
The observed neutrino oscillation data suggest the right-handed
 neutrino mass scale to be around $10^{13}-10^{14}$ GeV,
 which is a few orders of magnitude below the GUT scale.
With fixed Yukawa couplings of right-handed neutrinos
 in the minimal SO(10) model, this intermediate scale is provided
 by the vacuum expectation value (VEV) of
 the $\overline{\bf 126}$ Higgs multiplet.
This indicates the existence of many exotic states with intermediate
 mass scale, which significantly alter the running
 of the MSSM gauge couplings.
This has been discussed in Ref.~\cite{GCU},
 where it is shown that the gauge couplings are not unified any more,
 and even the SU(2) gauge coupling blows up below the $M_{GUT}$.
To solve this problem, we may extend the minimal model
 or may consider a different direction
 in constructing GUT models \cite{5DGUT}.

In this paper we consider a supersymmetric
 SU(4)$_c  \times$SU(2)$_L \times$SU(2)$_R$
 (4-2-2) model with a set of Higgs multiplets 
 which closely resembles the minimal SO(10) model.
The Higgs multiplets which play an important role in our discussion are 
 $H_{1,2,2}: ({\bf 1},{\bf 2},{\bf 2})$, 
 $H_{15,2,2}: ({\bf 15},{\bf 2},{\bf 2})$,     
 $H_{10,1,3}: ({\bf 10},{\bf 1},{\bf 3})$, and $\overline{H}_{10,1,3}: (\overline{\bf 10},{\bf 1},{\bf 3})$ 
 corresponding to the Higgs multiplets in the minimal SO(10) model, namely
 $H_{1,2,2} \subset$ {\bf 10}-plet Higgs and
 $H_{15,2,2} + H_{10,1,3} \subset \overline{\bf 126}$-plet Higgs. 
In the SO(10) model context, the Higgs multiplet  $\overline{H}_{10,1,3}$ belongs 
  to ${\bf 126}$-plet Higgs, which is introduced to satisfy the $D$-flat condition
  along with the $\overline{\bf 126}$-plet Higgs. 
The SU(4)$_c  \times$SU(2)$_L \times$SU(2)$_R$ symmetry is broken 
 down to the MSSM gauge group by 
  VEVs of $ \langle H_{10,1,3} \rangle =\langle \overline{H}_{10,1,3} \rangle$ 
  satisfying the $D$-flat condition. 
Although we are not going to details of the Higgs potential,  
  it is easy to realize a successful Higgs sector of our model  
  by analogy to the minimal set of Higgs multiplets in the minimal renormalizable 
  SO(10) model \cite{minimalHiggs}. 
In the SO(10) model, the minimal set of Higgs multiplets consists of ${\bf 10}$-plet, ${\bf 126}$-plet, 
  $\overline{{\bf 126}}$-plet and ${\bf 210}$-plet.  
It has been demonstrated in Ref.~\cite{minimalHiggs} that the most general renormlizable superpotential 
  for the minimal set of Higgs multiplets realizes the SO(10) symmetry breaking to the MSSM gauge group,  
  leaving only the MSSM particle contents light. 
Since our 4-2-2 model can be embedded in the minimal renormalizable SO(10) model, 
  we can consider a successful Higgs sector of our model as a subset of the Higgs sector 
  of the minimal SO(10) model.

The superpotential relevant for the fermion mass matrices
 is given by
\bea
 W = Y_{1}^{ij}  F_i {\bar F}_j H_{1,2,2} +
     Y_{15}^{ij} F_i {\bar F}_j H_{15,2,2} +
     Y_R^{ij}   {\bar F}_i {\bar F}_j H_{10,1,3},
\label{Yukawa1}
\eea
where $F_i: ({\bf 4}, {\bf 2}, {\bf 1})$ and
 ${\bar F}_i: (\overline{\bf 4}, {\bf 1}, {\bf 2})$
 denote the matter multiplets in $i$-th generation ($i=1,2,3$).
Assuming appropriate VEVs for the Higgs multiplets,
 we can parameterize the fermion mass matrices as the follows:
\bea
  M_u &=& c_1 M_{1,2,2} +  c_{15} M_{15,2,2} \; , \nonumber \\
  M_d &=&     M_{1,2,2} +         M_{15,2,2} \; , \nonumber \\
  M_D &=& c_1 M_{1,2,2} -3 c_{15} M_{15,2,2} \; , \nonumber \\
  M_e &=&     M_{1,2,2} -3        M_{15,2,2} \; , \nonumber \\
  M_R &=&     M_{10,1,3} .
\eea
 Here $M_u, M_d,$ are the mass matrices for up-type and down-type quarks,
 $M_D$ is the neutrino Dirac mass matrix,
 $M_e$ is the charged lepton mass matrix,
 and $M_R$ is right-handed Majorana neutrino mass matrix.
They are given in terms of the three fundamental matrices
 $M_{1,2,2}$, $M_{15,2,2}$ and $M_{10,1,3}$
 and the complex coefficients $c_1$ and $c_{15}$.
Note that the relations between fermion mass matrices
 are exactly the same as those in the minimal SO(10) model,
 except for $M_R$.
In the 4-2-2 model, $M_R$ is independent of the other mass matrices,
 while it is proportional to $M_{15,2,2}$ in the minimal SO(10) model.

As is well-known, the MSSM gauge couplings successfully unify 
  at $M_{GUT} \simeq 2 \times 10^{16}$ GeV. 
In the minimal SO(10) model, $M_{GUT}$ is the scale at which the SO(10) gauge symmetry 
  is broken down to the MSSM gauge symmetry. 
Since the 4-2-2 model with left-right symmetry\footnote{
The left-right symmetry requires us to add
 $Y_R^{ij} F_i F_j H_{10,3,1}$ to Eq.~(\ref{Yukawa1})
 with a Higgs multiplet
 $H_{10,3,1}: (\overline{\bf 10},{\bf 3},{\bf 1})$.
This term corresponds to type II seesaw \cite{SeesawII}
 once $H_{10,3,1}$ develops a non-zero VEV.
Since a more complicated Higgs sector seems necessary
 to induce such a VEV, we do not consider type II seesaw
 in this paper.
} 
can be embedded in the SO(10) model, 
  we simplify identify $M_{GUT}$ with the breaking scale of 4-2-2 down to the MSSM gauge group, 
  assuming the left-right symmetry. 
Therefore, the procedure for fitting the charged fermion
 mass matrices is the same as in the minimal SO(10) model.
On the other hand, it is important to note that $M_R$ being independent of the other mass matrices
 provides us with the freedom to fit the neutrino oscillation data.

Let us count here the number of free parameters
 used to fit the charged fermion mass matrices.
Because of left-right symmetry,
 $M_{1,2,2}$ and $M_{15,2,2}$ are $3\times3$ complex symmetric matrices.
Without loss of generality, we take a basis
 where $M_{1,2,2}$ is real and diagonal,
 so that the number of free parameters
 in $M_{1,2,2}$ and $M_{15,2,2}$ is $3+12=15$.
The two complex parameters $c_1$ and $c_{15}$ introduce 
 an additional 4 degrees of freedom, and therefore 
 in total we have 19 free parameters.
The degrees of freedom of charged fermion mass matrices
 are decomposed into $3+6=9$ for the lepton and quark
 mass eigenvalues, and another $9$ for a unitary matrix
 for the quark mixings which consists of 4 parameters in
 the CKM matrix and 5 diagonal CP-phases.
Since the 5 CP-phases are not observable in the SM,
 we drop these degrees of freedom.
Thus, we have 14 free parameters to fit 13 observables
 \cite{MatsudaEtal}.
In the minimal SO(10) model, this single free parameter is adjusted
 to fit the neutrino oscillation data
 (see \cite{Fukuyama-Okada} for details).

Through the type I seesaw mechanism \cite{Seesaw},
 the light neutrino mass matrix is given by
\bea
  m_\nu = Y_D^T M_R^{-1} Y_D v_u^2 = U_{PMNS}^* D_\nu U_{PMNS}^\dagger,
\label{eq3}
\eea
where $Y_D$ is the neutrino Dirac Yukawa matrix and $v_u$ is the VEV of the up-type Higgs doublet.
The PMNS mixing matrix, by which $m_\nu$ is diagonalized to the mass eigenvalue matrix $D_\nu$, 
 is parametrized as 
\bea
U_{PMNS} = \begin{bmatrix} c_{12} c_{13}&c_{12}c_{13}&s_{13}e^{-i\delta_{CP}}\\-s_{12}c_{23}-c_{12}s_{23}s_{13}e^{i\delta_{CP}}&c_{12}c_{23}-s_{12}s_{23}s_{13}e^{i\delta_{CP}}&s_{23} c_{13}\\ s_{12}c_{23}-c_{12}c_{23}s_{13}e^{i\delta_{CP}}&-c_{12}s_{23}-s_{12}c_{23}s_{13}e^{i\delta_{CP}}&c_{23}c_{13}
\end{bmatrix} 
\begin{bmatrix}
1&0&0\\
0&e^{-i \rho_1}&0\\
0&0&e^{-i \rho_2}
 \end{bmatrix},
\eea
where $c_{ij}=\cos\theta_{ij}$,  $s_{ij}=\sin\theta_{ij}$, and $\rho_1$ and $\rho_2$ are the Majorana phases. 
Using Eq.~(\ref{eq3}), we can express the right-handed neutrino
 mass matrix as
\bea
 M_R = v_u^2 \left(
  Y_D U_{MNS} D_\nu^{-1}U_{MNS}^T Y_D^T  \right).
\label{MRformula}
\eea
Recall that in this 4-2-2 model, $Y_D$ is fixed by fitting
  the Dirac fermion masses and mixings in the same manner as the minimal SO(10) model.
Hence, employing the current neutrino oscillation data (two neutrino mass squared differences 
  and three mixing angles), we obtain $M_R$  from Eq.~(\ref{MRformula}) as a function of 
  the lightest light neutrino mass eigenvalue, $\delta_{CP}$ and $\rho_{1, 2}$.

Models with the seesaw mechanism can also
 account for generating the observed baryon asymmetry
 in the universe \cite{Planck2015},
\bea
 Y_B = \frac{n_B}{s} = (8.6-9.0) \times 10^{-11}
 \label{BAU_data}
\eea
 via thermal leptogenesis \cite{Fukugita-Yanagida},
 where $Y_B$ is the ratio of the baryon (minus anti-baryon) density
 ($n_B$) to the entropy density ($s$).
The out-of-equilibrium decays of heavy Majorana neutrinos
 in the presence of non-zero CP-violating phase generates
 a lepton asymmetry $Y_L$ in the universe, which is partially
 converted to the baryon asymmetry through (B+L)-violating sphaleron
 transitions \cite{sphaleron1, sphaleron2}.
The conversion rate is given by \cite{conversion}
\bea
 Y_B = - \frac{8 N_f + 4N_H}{22 N_f + 13 N_H} Y_L
     = - \frac{8}{23} Y_L.
\label{rate}
\eea
Here we set  $N_f = 3$ and $N_H = 2$ for the numbers
 of fermion families $N_f$ and Higgs doublets $N_H$
 as in the minimal SUSY SM (MSSM).

The baryon asymmetry produced is evaluated by solving the Boltzmann equations with the information of
 neutrino Dirac Yukawa coupling matrix and $M_R$.
Since $Y_D$ is fixed and $M_R$ is a function of $\delta_{CP}$ and $\rho_{1,2}$ in our model,
 the resultant baryon asymmetry is given as a function of these parameters. 
Therefore, leptogenesis constrains the parameters, $\delta_{CP}$ and $\rho_{1,2}$, 
 so as to reproduce the observed baryon asymmetry.

As mentioned above, the data fitting procedure
 for the realistic charged fermion mass matrices
 is the same as in the minimal SO(10) model,
 and so in our analysis we employ the numerical values
 in $Y_D$ found in \cite{Fukuyama-Okada}.
In the basis where the charged lepton mass matrix is diagonal,
 the neutrino Dirac Yukawa coupling matrix at the GUT scale
 is unambiguously determined and explicitly given by
\begin{eqnarray}
 Y_D =
\left(
 \begin{array}{ccc}
-0.000135 - 0.00273 i & 0.00113  + 0.0136 i  & 0.0339   + 0.0580 i  \\
 0.00759  + 0.0119 i  & -0.0270   - 0.00419  i  & -0.272    - 0.175   i  \\
-0.0280   + 0.00397 i & 0.0635   - 0.0119 i  &  0.491  - 0.526 i
 \end{array}   \right) ,
\label{Ynu1}
\end{eqnarray}
 for $\tan \beta=45$.\footnote{
Although the output for the neutrino oscillation parameters
 obtained in \cite{Fukuyama-Okada} is more than 3$\sigma$ away from
 the current neutrino oscillation data \cite{NuData},
 the experimental data for charged fermion mass matrices are nicely fitted.
Since $Y_D$ is determined only by data-fitting the charged fermion mass matrices,
 we can safely use this $Y_D$ data without contradicting any of the experimental results.
}
\footnote{
In Ref.~\cite{Fukuyama:2003hn}, the charged lepton flavor violating (LVF) processes have been 
  investigated in the minimal SO(10) model with the $Y_D$ data of Eq.~(\ref{Ynu1}). 
Although the right-handed neutrino mass spectrum in our 4-2-2 model is different from the one 
  in the minimal SO(10) model, we expect that the rate of the LFV processes lies in the same order 
  to those presented in \cite{Fukuyama:2003hn}.  
Considering the final results of the MEG experiment \cite{Mori:2016vwi}, 
  ${\rm BR}( \mu^+ \to e^+ \gamma) < 4.2 \times 10^{-13}$ at 90\% C.L., 
  we see from the results in \cite{Fukuyama:2003hn} that 
  the lower mass bounds on sleptons and winos will be multi-TeV 
  to avoid the MEG constraint. 
}
Using $Y_D$, we determine $M_R$ from Eq.~(\ref{MRformula})
 as a function of $\delta_{CP}$ and $\rho_{1,2}$.
Since the absolute mass spectrum of light neutrinos has not
 yet been determined, we consider two cases for it,
 the normal hierarchical (NH) case and
 the inverted hierarchical case (IH).
For the NH case, the mass eigenvalue matrix $D_\nu$ is given by
\bea
 D_\nu=\mbox{diag}\left(
m_0,
~\sqrt{m_0^2 + \Delta m_{12}^2},
~\sqrt{m_0^2 + \Delta m_{12}^2 + \Delta m_{23}^2}\right),
\label{Dnu-NH}
\eea
while for the IH case
\bea
 D_\nu=\mbox{diag}
 \left(
\sqrt{m_0^2  - \Delta m_{12}^2 + \Delta m_{23}^2},
~\sqrt{m_0^2 + \Delta m_{23}^2},
~m_0 \right),
\label{Dnu-IH}
\eea
 where the lightest mass eigenvalue $m_0$
 is a free parameter.
In our analysis, we adopt the following values
 for the neutrino oscillation data \cite{NuData, An:2012eh}:
\bea
&& \Delta m_{12}^2=7.6\times 10^{-5}~\mbox{eV}^2,
 ~~\Delta m_{23}^2=2.4\times 10^{-3}~\mbox{eV}^2 , \nonumber \\
&& \sin^2 (2  \theta_{12})=0.87, ~~\sin^2 (2  \theta_{23})=1.0 , ~~\sin^2 (2  \theta_{13})=0.092 . 
\label{nudata}
\eea

\begin{figure}[t]
\begin{tabular}{cc}
\begin{minipage}{0.5\hsize}
\begin{center}
{\includegraphics[scale=.9]{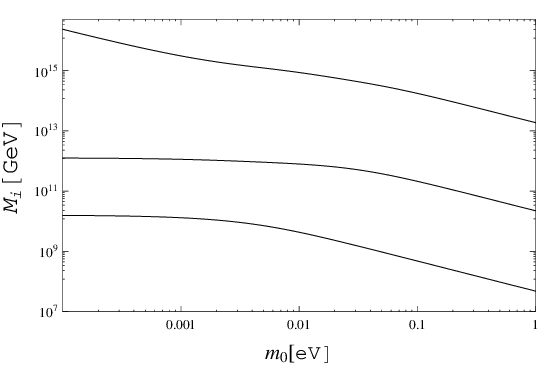}}
\end{center}
\end{minipage}
\begin{minipage}{0.5\hsize}
\begin{center}
{\includegraphics[scale=.9]{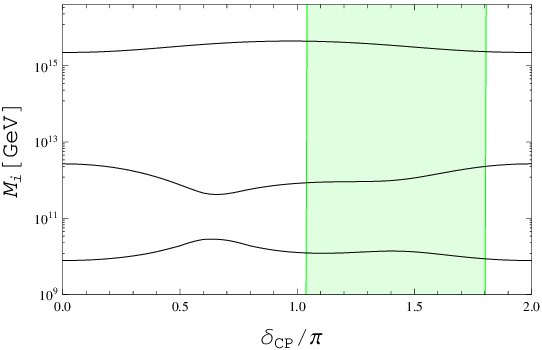}}
\end{center}
\end{minipage}
\end{tabular}
\caption{
For the NH case, heavy neutrino mass spectrum versus $m_0$ for $\delta_{CP}=\frac{3 \pi}{2}$ (left panel),
 and versus $\delta_{CP}$ for $m_0=10^{-3}$ eV (right panel).
The (green) shaded region denotes the allowed region for $\delta_{CP}$ at the 95\% confidence level 
 by the recent T2K data \cite{T2K}. 
}
\label{fig:massNH}
\end{figure}

\begin{figure}[t]
\begin{tabular}{cc}
\begin{minipage}{0.5\hsize}
\begin{center}
{\includegraphics[scale=.9]{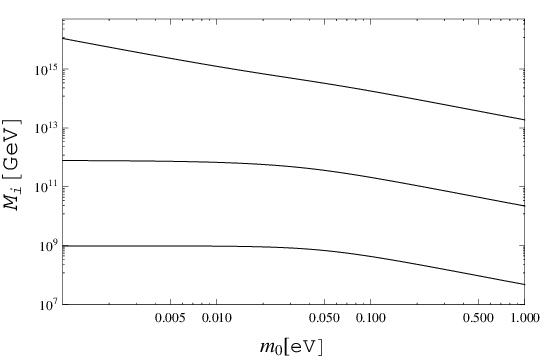}}
\end{center}
\end{minipage}
\begin{minipage}{0.5\hsize}
\begin{center}
{\includegraphics[scale=.9]{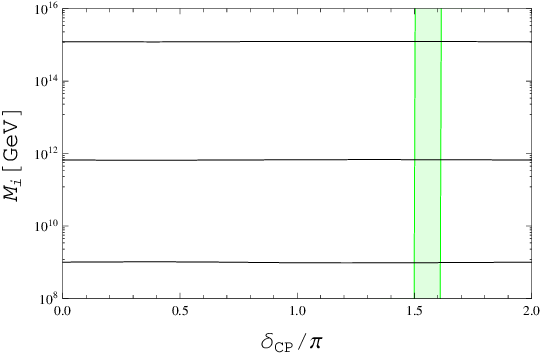}}
\end{center}
\end{minipage}
\end{tabular}
\caption{
Same as Figure \ref{fig:massNH} but for the IH case. 
}
\label{fig:massIH}
\end{figure}

Let us first show the mass spectrum of the heavy Majorana neutrinos
  (mass eigenvalues of $M_R$) as a function of $m_0$ and $\delta_{CP}$. 
For simplicity, we set $\rho_{1,2}=0$ here. 
Figure~\ref{fig:massNH} (left panel) shows the mass spectrum $M_i$  $(i=1,2,3)$
 of the heavy Majorana neutrinos for the NH case
 as a function of $m_0$ with $\delta_{CP}=3 \pi/2$ indicated by the recent T2K and NO$\nu$A data.   
Since the VEV of $H_{10,1,3}$ (which breaks 4-2-2 down to MSSM)
 is $M_{GUT}$, we require $m_0 \gtrsim 10^{-4}$ eV in order to
 keep $Y_R^{ij}$ within the perturbative regime.
The right panel shows the mass spectrum as a function of $\delta_{CP}$ for $m_0=10^{-3}$ eV. 
The (green) shaded region denotes the allowed region for $\delta_{CP}$ at the 95\% confidence level by the recent T2K data \cite{T2K}. 
The corresponding results for the IH case are shown in Figure~\ref{fig:massIH}.
For the IH case, we find $M_1 \lesssim 10^9$ GeV for any values of $m_0$ and $\delta_{CP}$.

As shown in Figures~\ref{fig:massNH} and \ref{fig:massIH},
 the heavy neutrino masses are hierarchical
 for both the NH and IH cases.
The lepton asymmetry in the universe in this case is dominantly
 produced by the lightest heavy neutrino decay,
 since the asymmetry produced by heavier neutrino decays
 are almost completely washed-out \cite{Plumacher1}.
Thus, we consider the lepton asymmetry produced by only
 the lightest heavy neutrino decay.
In addition, there is a lower bound on the lightest heavy
 neutrino mass, $M_1 \gtrsim 10^{9-10}$ GeV, to produce
 the desired amount of baryon asymmetry \cite{bound}.
For the IH case, the lightest heavy neutrino mass
 is always found to be below this bound, and
 in our numerical analysis we find that the
 resultant baryon asymmetry is too small in comparison
 to the observed baryon asymmetry.
Therefore, in the following, we present our results only for the NH case.

For a successful thermal leptogenesis, the reheating temperature ($T_r$) 
  after inflation must be higher than $M_1$ for the lightest heavy neutrino 
  to be in thermal equilibrium at $T_r$.  
The heavy neutrino mass spectrum shown in Figure~\ref{fig:massNH} 
  indicates a lower bound on $T_r  > 10^9-10^{10}$ GeV.  
On the other hand, in SUSY scenarios there is an upper bound on reheating temperature 
  from the cosmological gravitino problem. 
According to the analysis in Ref.~\cite{Kawasaki:2008qe}, 
  we find $T_r < 10^6-10^{10}$ TeV depending on the sparticle mass spectrum, 
  in particular with a gravitino mass in the range of $m_{3/2}=1-10$ TeV. 
In our scenario, we assume the gravitino mass of order 10 TeV or higher, 
  so that the reheating temperature can be higher than $M_1$
  while avoiding the cosmological gravitino problem.

Since our model is supersymmetric, we need to consider
 the lepton asymmetry generated by the decays of
 both the lightest heavy neutrino and sneutrino.
From Figure~\ref{fig:massNH},
 the lightest heavy neutrino mass
 is far below $M_{GUT}$, and so the effective theory
 for leptogenesis contains the MSSM and three light neutrinos,
 as well as the lightest heavy neutrino superfield.
Although the complete Boltzmann equations for this system
 is quite involved (see \cite{Plumacher2} for complete formulas),
 because of supersymmetry the lepton asymmetry stored
 in the SM particles is exactly the same as that stored
 in the sparticles \cite{Plumacher2}.
Since the heavy neutrino mass scale is much higher than
 the typical sparticle mass scale $\sim$ TeV,
 our system is supersymmetric to a very good approximation.
Among the many decay and scattering processes involved
 in the Boltzmann equations, it is known that
 the (inverse) decay process of the lightest heavy (s)neutrino
 plays the most important role in determining
 the resultant baryon asymmetry, while the others are negligible
 in most of the parameter space \cite{Plumacher1}.
Including only the decay process greatly simplifies
 the Boltzmann equations, so that for the heavy neutrino
 they are exactly the same as  in the non-supersymmetric case:
\bea
 \frac{dY_{N_1}}{dz} &=& \frac{-z}{s H(M_1)}
 \left(\frac{Y_{N_1}}{Y_{N_1}^{eq}}-1 \right) \gamma_{N_1}, \nonumber \\
 \frac{dY_{L_f}}{dz}&=&-\frac{z}{sH(M_1)}
 \left[\frac{1}{2} \frac{Y_{L_f}}{Y_l^{eq}}
 +\epsilon_1\left(\frac{Y_{N_1}}{Y_{N_1}^{eq}}-1\right)\right]
  \gamma_{N_1},
\label{Boltzmann}
\eea
where $Y_{N_1}$ is the yield (the ratio of the number density to
 the entropy density $s$) of the lightest heavy neutrino,
 $Y_{N_1}^{eq}$ is the yield in thermal equilibrium,
 the  temperature of the universe is normalized by the mass of
 the heavy neutrino $z=M_1/T$,
 $H(M_1)$ is the Hubble parameter at $T=M_1$,
 and $Y_{L_f}$ is lepton asymmetry stored in the SM particles.
The $CP$-asymmetry parameter, $\epsilon_1$, is given by \cite{epsilon}
\bea
\epsilon_1 = - \frac{1}{2 \pi (Y_\nu Y_\nu^\dag)_{11}}
 \sum_{j\neq 1} \mbox{Im} \left[(Y_\nu Y_\nu^\dag)_{1j}^2 \right]
 f(M_j^2/M_1^2),
\label{epsilon}
\eea
where
\bea
f(x) &\equiv&
 \sqrt{x} \; \mbox{ln}\left(\frac{1+x}{x}\right)
+ 2 \frac{\sqrt{x}}{x - 1},
\eea
and $Y_\nu$ is the neutrino Dirac Yukawa coupling matrix
 in the basis where both the charged lepton matrix and
 $M_R$ are diagonalized.
Using Eqs.~(\ref{MRformula}) and (\ref{Ynu1}),
 we can obtain $Y_\nu$ as a function of $m_0$, $\delta_{CP}$ and $\rho_{1,2}$.

The space-time density of the heavy neutrino decay
  in thermal equilibrium, $\gamma_{N_1}$ is given by
\bea
 \gamma_{N_1} = s Y_{N_1}^{eq} \frac{K_1(z)}{K_2(z)} \Gamma_{N_1},
\eea
 where $K_1$ and $K_2$ are the modified Bessel functions,
 and
\bea
\Gamma_{N_1}= \frac{(Y_\nu Y_\nu^\dagger)_{11}}{8 \pi} M_1
\eea
 is the decay width of the heavy neutrino.
Then, we solve the Boltzmann equations
 with the boundary conditions
 $Y_{N1}(0) = Y_{N_1}^{eq}(0)$ and $Y_{L_f}(0) = 0$.
The lepton asymmetry generated by the right-handed neutrino decays
 is converted to the baryon asymmetry via the sphaleron process
 with the rate of Eq.~(\ref{rate}) and hence,
 we evaluate the resultant baryon number as
\bea
 Y_B = - \frac{8}{23} Y_{L_f}(\infty) \times 2,
\eea
 where the factor 2 takes into account
 the baryon number stored in sparticles.

\begin{figure}[t]
\begin{center}
\epsfig{file=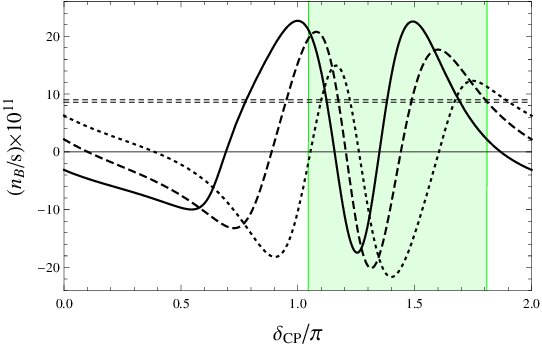, width=12cm}
\caption{
Baryon asymmetry as a function of $\delta_{CP}$ for $m_0=10^{-3}$ eV, $\rho_2=0$ 
  and $\rho_{1}=0$ (solid), $\frac{\pi}{6}$ (dashed) and $\frac{\pi}{3}$ (dotted). 
The dashed horizontal lines show the range of the observed baryon asymmetry in Eq.~(\ref{BAU_data}).
The (green) shaded region denotes the allowed region for $\delta_{CP}$ at the 95\% confidence level 
 by the recent T2K data \cite{T2K}. 
}
\label{fig:BAU1}
\end{center}
\end{figure}

\begin{figure}
\begin{center}
\epsfig{file=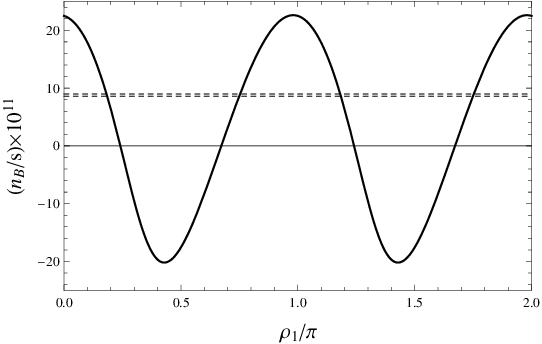, width=12cm}
\caption{
Baryon asymmetry as a function of a Majorana phase $\rho_1$
 for $m_0=10^{-3}$ eV, $\delta_{CP}=\frac{3 \pi}{2}$ and $\rho_2=0$.  
The dashed horizontal lines show the range of the observed baryon asymmetry in Eq.~(\ref{BAU_data}).
 }
\label{fig:BAU2}
\end{center}
\end{figure}

For various values of the free parameters ($m_0$, $\delta_{CP}$ and $\rho_{1,2}$), 
  we numerically solve the Boltzmann equations.
In our analysis, we fix $m_0=10^{-3}$ eV and $\rho_2=0$, for simplicity. 
For $m_0 \lesssim 10^{-2}$ eV, $M_1$ is almost independent
 of $m_0$, and we find that the results
 for the generated baryon asymmetry  are almost the same.
Figure~\ref{fig:BAU1} shows the resultant baryon asymmetries
 as a function of $\delta_{CP}$ for three different values of 
 the Majorana phase, namely, $\rho_1=0$ (solid), $\frac{\pi}{6}$ (dashed) and $\frac{\pi}{3}$ (dotted), 
 along with the observed value (horizontal lines). 
The allowed region for $\delta_{CP}$ at the 95\% confidence level 
  from the recent T2K data is depicted by the (green) shaded region. 
We have found the parameters in the shaded region to reproduce the observed baryon asymmetry. 
Figure~\ref{fig:BAU2} shows the baryon asymmetries
  as a function of $\rho_1$ for $\delta_{CP}=\frac{3 \pi}{2}$, 
  along with the observed value (horizontal lines). 
A suitable choice of $\rho_1$ can reproduce the observed baryon asymmetry.

In summary, we have considered a supersymmetric
 SU(4)$_c  \times$SU(2)$_L \times$SU(2)$_R$ model
  with a minimal number of Higgs multiplets and $CP$-violating phases ($\delta_{CP}$ and $\rho_{1,2}$) 
  in the neutrino flavor mixing matrix.
The model has the same structure in the Yukawa couplings
  for the charged fermions as the supersymmetric minimal SO(10) model,
  so that the neutrino Dirac Yukawa coupling matrix is unambiguously determined 
  by fitting the experimental data for charged fermion mass matrices.
Using the type I seesaw formula with the neutrino Dirac Yukawa coupling matrix, 
  the right-handed Majorana neutrino mass matrix is given as a function of
  $m_0$, $\delta_{CP}$ and $\rho_{1,2}$.
We have employed leptogenesis and the observed baryon asymmetry to identify 
  the allowed parameter regions. 
Only the NH case for the light neutrino mass spectrum can reproduce the observed baryon asymmetry 
  with a suitable choice of $\delta_{CP}$ and $\rho_{1,2}$.
We have found that the  Dirac $CP$-violating phase around $\delta_{CP}=\frac{3 \pi}{2}$, 
  which is strongly indicated by the recent T2K and NO$\nu$A data, 
  leads to the baryon asymmetry compatible to the observed value.  
Once $\delta_{CP}$ has been more precisely determined, 
  the allowed regions for $\rho_{1,2}$ will be determined.

\section*{Acknowledgments}
N.O.~would like to thank the Particle Theory Group
 of the University of Delaware for hospitality during his visit.
This work is supported in part by the DOE Grants: 
 No.~DE-SC0012447(N.O.) and No.~DE-SC0013880 (Q.S.).




\begin{thebibliography}{99}
\bibitem{T2K}
  K.~Abe {\it et al.} [T2K Collaboration],
  Phys.\ Rev.\ D {\bf 96}, no. 9, 092006 (2017); 
https://www.mpi-hd.mpg.de/nu2018/programme.


\bibitem{NOVA} 
  P.~Adamson {\it et al.} [NOvA Collaboration],
  Phys.\ Rev.\ Lett.\  {\bf 118}, no. 23, 231801 (2017); 
https://www.mpi-hd.mpg.de/nu2018/programme.

\bibitem{Seesaw}
P.~Minkowski, Phys. Lett. B {\bf 67}, 421 (1977);
T.~Yanagida, in \emph{Proceedings of the Workshop on the Unified
  Theory and the Baryon Number in the Universe} (O.~Sawada and
  A.~Sugamoto, eds.), KEK, Tsukuba, Japan, 1979, p.~95;
M.~Gell-Mann, P.~Ramond, and R.~Slansky, \emph{Supergravity} (P.~van
  Nieuwenhuizen et al. eds.), North Holland, Amsterdam, 1979, p.~315;
S.~L. Glashow, \emph{The future of elementary particle physics}, in
  \emph{Proceedings of the 1979 Carg{\`e}se Summer Institute
 on Quarks and Leptons} (M.~Levy et al. eds.),
 Plenum Press, New York, 1980, p.~687;
R.~N. Mohapatra and G.~Senjanovic,
 Phys. Rev. Lett. {\bf 44}, 912 (1980).


\bibitem{BM}
 K.~S.~Babu and R.~N.~Mohapatra,
 Phys.\ Rev.\ Lett.\  {\bf 70}, 2845 (1993).

\bibitem{Lazarides:1980nt}
  G.~Lazarides, Q.~Shafi and C.~Wetterich,
  Nucl.\ Phys.\  B {\bf 181}, 287 (1981);
  For an SO(10) model which also includes the ${\bf 120}$ representation see
  C.~Panagiotakopoulos, Q.~Shafi and C.~Wetterich,
  Phys.\ Rev.\ Lett.\  {\bf 55}, 787 (1985).

\bibitem{Ananthanarayan:1991xp}
  B.~Ananthanarayan, G.~Lazarides and Q.~Shafi,
  Phys.\ Rev.\  D {\bf 44}, 1613 (1991).


\bibitem{Fukuyama-Okada}
T.~Fukuyama and N.~Okada,
  JHEP {\bf 0211}, 011 (2002).


\bibitem{datafit}
  H.~S.~Goh, R.~N.~Mohapatra and S.~P.~Ng,
  Phys.\ Rev.\ D {\bf 68}, 115008 (2003);
  B.~Dutta, Y.~Mimura and R.~N.~Mohapatra,
  Phys.\ Rev.\ D {\bf 69}, 115014 (2004);
%
{\it ibid.},
  Phys.\ Lett.\ B {\bf 603}, 35 (2004);
%
  S.~Bertolini, M.~Frigerio and M.~Malinsky,
  Phys.\ Rev.\ D {\bf 70}, 095002 (2004);
%
  S.~Bertolini and M.~Malinsky,
  Phys.\ Rev.\ D {\bf 72}, 055021 (2005);
%
  K.~S.~Babu and C.~Macesanu,
  Phys.\ Rev.\ D {\bf 72}, 115003 (2005).

\bibitem{Mimura}
For recent analysis, see 
 T.~Fukuyama, K.~Ichikawa and Y.~Mimura,
  Phys.\ Rev.\ D {\bf 94}, no. 7, 075018 (2016). 


\bibitem{p-dcay}
  H.~S.~Goh, R.~N.~Mohapatra, S.~Nasri, and S.~-P.~Ng,
  Phys.\ Lett.\  {\bf B587}, 105-116 (2004); 
%
  T.~Fukuyama, A.~Ilakovac, T.~Kikuchi, S.~Meljanac, and N.~Okada,
  JHEP {\bf 0409}, 052 (2004).


\bibitem{LFV}
  T.~Fukuyama, T.~Kikuchi and N.~Okada,
  Phys.\ Rev.\  {\bf D68}, 033012 (2003);
  [arXiv:1104.1736 [hep-ph]].


\bibitem{GCU}
  S.~Bertolini, T.~Schwetz and M.~Malinsky,
  Phys.\ Rev.\ D {\bf 73}, 115012 (2006).


\bibitem{5DGUT}
T.~Fukuyama, T.~Kikuchi and N.~Okada,
  Phys.\ Rev.\  {\bf D75}, 075020 (2007); 
%
R.~N.~Mohapatra, N.~Okada and H.~-B.~Yu,
  Phys.\ Rev.\  {\bf D76}, 015013 (2007); 
%
T.~Fukuyama and N.~Okada,
  Phys.\ Rev.\  {\bf D78}, 015005 (2008); 
{\it ibid.}, Phys.\ Rev.\  {\bf D78}, 115011 (2008).


\bibitem{minimalHiggs}
C.~S.~Aulakh, B.~Bajc, A.~Melfo, G.~Senjanovic and F.~Vissani,
  Phys.\ Lett.\ B {\bf 588}, 196 (2004)
  [hep-ph/0306242]; 
%
T.~Fukuyama, A.~Ilakovac, T.~Kikuchi, S.~Meljanac and N.~Okada,
  Eur.\ Phys.\ J.\ C {\bf 42}, 191 (2005)
  [hep-ph/0401213].  


\bibitem{SeesawII}
G.~Lazarides, Q.~Shafi and C.~Wetterich,
 Nucl.\ Phys.\ {\bf B181}, 287 (1981);
R.~ N.~ Mohapatra and G.~Senjanovi\'c,
 Phys.\  Rev.\ {\bf D 23}, 165 (1981);
M.~Magg and C.~Wetterich, Phys.\ Lett.\  B {\bf 94}, 61 (1980);
J.~Schechter and J.~W.~F.~Valle, Phys.\ Rev.\  D {\bf 22}, 2227 (1980).



\bibitem{MatsudaEtal}
  K.~Matsuda, Y.~Koide, T.~Fukuyama, and H.~Nishiura,
  Phys.\ Rev.\  {\bf D65}, 033008 (2002).


\bibitem{Planck2015} 
  P.~A.~R.~Ade {\it et al.} [Planck Collaboration],
  Astron.\ Astrophys.\  {\bf 594}, A13 (2016)
  [arXiv:1502.01589 [astro-ph.CO]].
  
  
\bibitem{Fukugita-Yanagida}
  M.~Fukugita and T.~Yanagida,
  Phys.\ Lett.\  {\bf B174}, 45 (1986). 

\bibitem{sphaleron1}
N.~S.~Manton,
  Phys.\ Rev.\  D {\bf 28}, 2019 (1983);
%
F.~R.~Klinkhamer and N.~S.~Manton,
  Phys.\ Rev.\  D {\bf 30}, 2212 (1984).


\bibitem{sphaleron2}
V.~A.~Kuzmin, V.~A.~Rubakov and M.~E.~Shaposhnikov,
  Phys.\ Lett.\  B {\bf 155}, 36 (1985).


\bibitem{conversion}
S.~Y.~Khlebnikov and M.~E.~Shaposhnikov,
  Nucl.\ Phys.\  B {\bf 308} (1988) 885.


\bibitem{NuData}
%
%
  C.~Patrignani {\it et al.} [Particle Data Group],
  ``Review of Particle Physics,''
  Chin.\ Phys.\ C {\bf 40}, no. 10, 100001 (2016).

\bibitem{Fukuyama:2003hn} 
  T.~Fukuyama, T.~Kikuchi and N.~Okada,
  Phys.\ Rev.\ D {\bf 68}, 033012 (2003)
  [hep-ph/0304190].
  


\bibitem{Mori:2016vwi} 
  T.~Mori [MEG Collaboration],
  Nuovo Cim.\ C {\bf 39}, no. 4, 325 (2017)
  [arXiv:1606.08168 [hep-ex]].  
  



\bibitem{An:2012eh} 
  F.~P.~An {\it et al.} [Daya Bay Collaboration],
  ``Observation of electron-antineutrino disappearance at Daya Bay,''
  Phys.\ Rev.\ Lett.\  {\bf 108}, 171803 (2012). 



\bibitem{Plumacher1}
M.~Plumacher,
  Z.\ Phys.\  {\bf C74}, 549-559 (1997).


\bibitem{bound}
W.~Buchmuller, P.~Di Bari and M.~Plumacher,
  Nucl.\ Phys.\  B {\bf 643}, 367 (2002)
  [Erratum-ibid.\  B {\bf 793}, 362 (2008)];
%
{\it ibid.},  Annals Phys.\  {\bf 315}, 305 (2005).



\bibitem{Kawasaki:2008qe} 
  M.~Kawasaki, K.~Kohri, T.~Moroi and A.~Yotsuyanagi,
  Phys.\ Rev.\ D {\bf 78}, 065011 (2008)
  [arXiv:0804.3745 [hep-ph]].


\bibitem{Plumacher2}
M.~Plumacher,
  Nucl.\ Phys.\  {\bf B530}, 207-246 (1998). 


\bibitem{epsilon}
L.~Covi, E.~Roulet and F.~Vissani,
  Phys.\ Lett.\  B {\bf 384}, 169 (1996).


\end{thebibliography}
\end{document}